\documentclass{aa}
\input epsf.sty
\input psfig.sty
\begin{document}
\thesaurus{02.13.2;08.16.6;09.10.1}
\title{Acceleration and collimation of relativistic plasmas ejected
by fast rotators}
\titlerunning{Acceleration and collimation of relativistic plasmas}
\author{S.V. Bogovalov}
\authorrunning{S.Bogovalov}
\institute{Moscow Engineering Physics Institute, Kashirskoje Shosse 31,
Moscow, 115409, Russia. e-mail: bogoval@axpk40.mephi.ru}
\date{Received 15 December 1999/Accepted 16 January 2001}
\maketitle
\begin{abstract}
A stationary self-consistent outflow of a magnetised relativistic plasma
from a rotating object with an initially monopole-like  magnetic field is
investigated in the ideal MHD approximation under the condition 
$\sigma/U_{0}^{2} > 1$,
where $\sigma$ is the ratio of the Poynting flux over the mass energy
flux at the equator and the surface of the star, with 
$U_0=\gamma_{0}v_0/c$ and $\gamma_0$ the initial four-velocity and 
Lorentz factor of the plasma. The mechanism of the magnetocentrifugal
acceleration and self-collimation of the relativistic plasma is 
investigated.  A jet-like relativistic flow along the axis
of rotation is found in the steady-state solution under the condition
$\sigma/U_{0}^{2} > 1 $ with properties predicted analytically. 
The amount of the collimated matter in the jet is rather small in 
comparison to the total mass flux in the wind. 
An explanation for the weak self-collimation of relativistic winds 
is given.
\end{abstract}
\section{Introduction}
Radio pulsars (Michel \cite{michel91}), 
superluminal galactic objects (Mirabel \& Rodriguez \cite{mirabel94}, 
\cite{mirabel99}) and AGNs (Begelman,
Blandford \& Rees \cite{blandford},
Pelletier et al. \cite{pelletier}) are observed to be associated with 
relativistic outflows. 
AGNs and superluminal galactic objects produce jets, while radio pulsars 
demonstrate 
a surprisingly high efficiency of transforming the pulsar rotational energy 
into kinetic energy of the relativistic wind
(Kennel \& Coroniti \cite{kennel}).
Hence, the  problem of acceleration and collimation of 
relativistic plasmas  is among the most important in the physics of 
relativistic outflows.

Nevertheless, radio pulsars, AGNs  and superluminal galactic objects 
are objects of rather different nature and scales.
It may well be that the mechanisms of acceleration of the outflows
in these sources are not the same. However, there seems to exist a mechanism 
which operates very well for all these objects, namely, the
so-called  magnetocentrifugal acceleration where the acceleration of the 
plasma outflow occurs in the magnetic field of the rotating object. 
Such conditions are realized in the magnetospheres of radio  pulsars,  in
the winds from accretion disks (Blandford \& Payne \cite{bp}) and apparently in  
superluminal galactic objects (Mirabel \& Rodriguez \cite{mirabel99}).
However, it has not been investigated whether the magnetocentrifugal 
mechanism could provide the acceleration
of the relativistic plasma in these objects to the observed bulk motion
velocities.

The analysis of problem of the relativistic plasma acceleration and collimation in a  
rotating magnetosphere started with Michel 
(\cite{michel}), where it was shown that the magnetocentrifugal
acceleration of the plasma is not effective for a prescribed monopole-like 
magnetic field.
Subsequently, the relativistic plasma outflow has been investigated in 
several other studies and for different astrophysical conditions
(Okamoto \cite{okamoto78}, Ardavan \cite{ardavan}, Beskin et al. \cite{bgi},
Camenzind \cite{max86}).
In particular, the ultrarelativistic plasma
outflow from an accretion disk was investigated numerically
in Camenzind (\cite{max87}) while the structure of electromagnetically driven
relativistic jets was discussed in Li (\cite{li93}) under the
assumption of jet confinement by an ambient medium.  Collimation of
stellar magnetospheres to jets in the force-free approximation was
studied in Sulkanen \& Lovelace (\cite{ml}) and 
Fendt et al. (\cite{fendt95}). Later, Fendt \& Camenzind (\cite{fendt96})
investigated  the plasma dynamics in the MHD approximation
in magnetic configurations obtained via the force-free approximation. 
A class of self-similar solutions of the 
relativistic plasma outflow has been obtained in Contopoulos (\cite{ianos94}). 
It was found that the efficiency of the plasma acceleration can be higher 
than in Michel's solution due to the effect of the magnetic collimation of
the wind to the axis of rotation (Begelman \& Li (\cite{li}),
Fendt \& Camenzind \cite{fendt96},
Takahashi \& Shibata (\cite {ts})). Thus, the problem of the
relativistic plasma acceleration in a rotating magnetic field
appeared to be closely connected to the problem of plasma collimation.

Strong evidence has been accumulated for a correlation between collimated astrophysical outflows 
(jets) and accretion disks. Direct evidence for a disk-jet
connection has been demonstrated by HST observations of the Herbig-Haro object
HH30 (Burrows et al. \cite{burrows}), although on the theoretical front  this
connection was discussed earlier (Camenzind \cite{max90}, Ferreira
\& Pelletier \cite{ferreira},
Falke \& Biermann \cite{falke}).  A disk-jet connection is also supported  by
observations of the active galaxy M87 (Ford et al. \cite{ford}) and
superluminal galactic objects (Mirabel \& Rodriguez \cite{mirabel98}).
These observations seem to lead to the conclusion that
jets require the existence of accretion disks for their formation
(e.g. Livio \cite{livio}) and  therefore accretion disks
should be a necessary element of the collimation mechanism.
However,  we have by now reliable evidence
that there is at least one source producing jets without 
the presence of an accretion disk. Namely, recent observations of the Crab
Nebula by the Chandra observatory leave no doubts that the Crab pulsar
indeed produces jets (Weiskopff et al. \cite{crab}).
Therefore, it would be important to understand the mechanism of plasma
collimation in all known jet sources, including the Crab pulsar.

Such a mechanism is provided by magnetic collimation of outflows from
accretion disks and it was first suggested by Lovelace (\cite{lovelace}),
Blandford (\cite{blandford76}) and Bisnovatyi-Kogan \& Ruzmaikin
(\cite{bkgr}). Subsequently, it was shown  from an analysis of the full 
set of the MHD equations for
nonrelativistic (Heayvaerts \& Norman \cite{heyvaerts}) and relativistic
outflows (Chiueh et al. \cite{chiueh}, Bogovalov \cite{bog95}) that 
collimation is a rather general property of magnetised winds from a
rotating object. A magnetised fast-mode supersonic
wind inevitably will be  partially cylindrically collimated   along the 
axis of rotation at distances large  compared to all 
characteristic dimensions of the source, such as the size of the 
accretion disk, provided that: the flow is non-dissipative,
the magnetised central object is rotating, the total magnetic flux of
one polarity reaching infinity (open field lines) is finite, the polytropic 
index of the plasma exceeds unity and the pressure of  ambient matter is 
low enough not to be able to terminate the wind before the formation of the jet.
This result is independent of  the specific properties of the
central source. Collimation  occurs spontaneously due to the internal Lorentz 
force arising in the magnetised plasma which is ejected from the rotating source
and therefore can be regarded as a process of self-collimation (hereafter
SC process). Note that the assumption of a polytropic gas with a constant 
polytropic index $\gamma$ does not seem really 
restrictive, since several exact MHD solutions for collimated outflows  have 
been 
constructed for the case of a variable index $\gamma$, Sauty \& Tsinganos 
(1994), 
Vlahakis \& Tsinganos (1998).   

Although the SC process has been questioned (e.g., Okamoto \cite{okamoto99}),
a numerical investigation  of this process and the  
characteristics of jets produced via this mechanism 
in the cold (Bogovalov \& Tsinganos \cite{bogts}) or hot
(Tsinganos \& Bogovalov \cite{tsbog}, Sauty et al (1999))
nonrelativistic winds leave no doubt that this mechanism works.
This mechanism provides the formation of cylindrically
collimated outflows at a large distance from the central source
while collimation is accompanied by essentially more
efficient acceleration of the plasma at the equatorial region as compared
with the original study of Michel (\cite{michel}). 
Evidence for the focusing  of  MHD outflows  
from accretion disks to the axis of rotation
at distances of the order of the dimension of accretion disks
has been obtained in other studies as well, such as in  Ouyed \& Pudritz 
(\cite{op}),
Ustyugova et.al. (\cite{ustyug}) and Krasnopolskii et. al. (\cite{krasno}).

In the analysis of Bogovalov \& Tsinganos (1999), the relaxation method 
was used to obtain the solution of the steady state outflow problem, 
i.e., the time-dependent problem was solved in order to get a steady-state 
solution. This method was first employed by Washimi \& Shibata 
(\cite{shibata}) 
for outflows from stellar objects. Later this method was applied in Bogovalov
(1996) for the investigation of the SC process
in an initially monopole-like magnetic field.
It seems that this numerical method allows us
to solve accurately all problems connected with the critical surfaces
(Bogovalov \cite{bog97b}, Tsinganos et al. \cite{tsincr}).
% The method  has been recently successfully used by several groups for the
%investigation of disk-winds (Ouyed \& Pudrits \cite{op},
%Ustyugova et al. \cite{ustyug}, Krasnopolskii et al. \cite{krasno},
%Koide et al. \cite{kazunari}).

In our previous study (Bogovalov \cite{bog97a}), the relaxation method has been 
applied to the investigation of the relativistic plasma outflow and 
it was found that the relativistic plasma is very weakly
collimated and accelerated even if the 
Poynting flux dominates the matter energy flux at the equator.
However, an analytical study performed in this work has shown
that an efficient SC of the relativistic
plasma is still possible for very fast rotation of the central source. 
Recently it has been argued that a  
relativistic plasma is poorly accelerated even under this condition 
(Beskin et al. \cite{beskin98}). However, 
the SC process was totally neglected in this work.
Therefore, the question concerning
the efficiency of the acceleration and SC of
the relativistic winds from fast rotators remains open.

\section{Basic assumptions}

Outflows of magnetised plasmas are described by the system of the familiar 
nonlinear MHD equations. It is not surprising that this system
has many solutions corresponding to different physical conditions.
Then, a reasonable choice of the appropriate approximation and 
a sound model is crucially important to arrive at some physically meaningful 
solution.

To formulate the basic demands of the model let us consider the general
properties of winds and jets. According to observations, the length of the jets 
greatly exceeds all characteristic scales of the central source.
Therefore, for an investigation of the SC process and the
characteristics of the jets, the theory should allow us to obtain the solution on
scales greatly exceeding all sizes of the central source.

Observations of winds and jets from several classes of astrophysical objects
such as our Sun (Parker \cite{parker}), or, 
YSO's (Burrows et al. \cite{burrows}), pulsars (Kennel \& Coroniti
\cite{kennel}) and AGNs (Ferrari et al. \cite{ferrari}) show that
they are super-magnetosonic at relatively large distances from the source.
At the same time all winds must be subsonic near the central source
where the basic parameters of the outflows are controlled.
Therefore, the theory must describe flows of mixed type, containing
the small scale regions of a subsonic flow and the large scale
regions of the super magnetosonic flow. The scales of these regions may differ
by several orders of magnitude. These properties of the outflows dictate
the choice of the approximations and mathematical models.

\subsection{Approximations}

The force-free approximation has been widely used
for the analysis of axisymmetric relativistic outflows 
(Scharlemann \& Wagoner \cite{sw}, Mestel \& Wang
\cite{mw}, Beskin et al. \cite{bgi}, Lyubarskii \cite{lyubarskii},
Fendt et al. \cite{fendt95}). For simplification reasons, 
the axisymmetric stellar magnetosphere can be assumed initially to 
have a dipolar magnetic field (Contopoulos et al.\cite{kazanas}). 
However,  this approximation is not appropriate for the
investigation of the SC process for several reasons. For example, 
it describes only subsonic flows of plasma since the force-free
transfield (Grad-Shafranov) equation is elliptic everywhere
(Fendt et al. \cite{fendt95}).
Therefore, this approximation cannot be used for the investigation
of plasma outflow in the supersonic regions.
Moreover, the collimating and accelerating forces are equal to zero in
this approximation. For these reasons, the force-free approximation
gives ambiguous results.  For example,
the structure of the flow  of the relativistic plasma at large distances from
the source  obtained in Sulkanen \& Lovelace  \cite{ml} and in Fendt et al. 
\cite{fendt95} differs from the 
structure obtained in Contopoulos et al. (\cite{kazanas}). Although Sulkanen \& 
Lovelace (\cite{ml}) and Contopoulos et al. (\cite{kazanas})
used similar models, Contopoulos et al. (\cite{kazanas}) obtained a 
radially
outflowing plasma while Sulkanen \& Lovelace  (\cite{ml})   obtained
collimated jets.  Jets were obtained as well in Fendt et al. (\cite{fendt95}). 
Actually all these works are mathematically correct.
The difference between them is only in the boundary conditions at  
infinity or at the outer boundary of the integration domain.
These works clearly demonstrate that the flow of plasma at large
distances from the source in the force-free approximation basically 
depends on the boundary conditions at infinity. 
But there is no such dependence  in real supersonic flows.
This is the most important qualitative difference between the subsonic and
super fast magnetosonic  flows.

In the steady state subsonic flow the MHD signals propagate down and
upstream of the flow.
That is why the plasma flow  is defined by the forces
affecting the plasma at a given point, upstream and down stream of the flow.
This is not the case in the super fast magnetosonic flow. 
No MHD signals can propagate upstream of the
supersonic flow. The supersonic  flow is defined only by the
forces which affect the plasma  upstream and at the given point and does
not depend on the forces which affect the plasma
downstream of the flow. For this reason the supersonic flow
does not depend on the boundary conditions located
downstream of the flow (actually the situation is more complicated;
the flow should cross the fast mode critical surface, Bogovalov \cite{bog97b}).
The simplest approximation which allows us to describe the mixed type outflows
is the ideal MHD approximation. This approximation is used in this paper.

\subsection{Model}

However,  even the use of the MHD approximation does
not always allow us to investigate the SC process in the supersonic region.
There is a wide class of  self-similar solutions for the nonrelativistic
(Blandford \& Payne \cite{bp},  Vlahakis \& Tsinganos \cite{tv} e.g.) and
relativistic plasma (Contopoulos \cite{ianos94}). All these solutions 
terminate not so far from the source.
Independent of why this termination happens, the self-similar solutions cannot be used for
the verification  of the predictions of the SC theory since 
the central source has infinitely large size with infinite magnetic flux
in the upper hemisphere. At any distance from the
source, the
dimension of the source exceeds  the distance to the
source. However, we are interested  in the opposite case, when the
distance is large compared
with the dimension of the central source.
Therefore, we demand that the appropriate model should have finite geometrical
dimensions, finite magnetic flux of the open field lines
of every polarity and initially
uncollimated plasma outflow. The last condition is important for us since our
objective is to study the process which transforms the initially uncollimated
outflow into the outflow containing the jet. The word ``initially'' means that
if one of the conditions for the SC process is not fulfilled (for example if
the central source is not rotating or the magnetic field is absent) the wind
expands radially at large distances from the source. Actually, a lot of models 
satisfy all these conditions.

The SC is a general property of magnetised winds from
rotating sources. It must take place in a wind from any source provided
that the conditions specified by Heayvaerts \& Norman (\cite{heyvaerts}), 
Chiueh et al. (\cite{chiueh}) and  Bogovalov (\cite{bog95}) 
are fulfilled. Therefore ``convenience'' and ``simplicity''
become reasonable  arguments for the choice of the
model for the investigation of this process.
The model of a star with a wind in an initially monopole-like (below m-l)
magnetic field firstly proposed by Sakurai (\cite{sakurai}) is the most
simple and convenient. This model allows us to avoid
the mathematically complicated problem connected
with the coexistence of open and closed field lines near the source
which actually has no direct relation to the SC process and usually is
considered particular (Keppens \& Goedbloed \cite{keppens}).

The model with the initially monopole-like magnetic field describes
more a wider class of outflows, including the time-dependent outflows from
oblique rotators (Bogovalov \cite{bog99}) and outflows from objects with
arbitrary distribution of the polarity of the magnetic field at the base
(Tsinganos \& Bogovalov \cite{tsbog}).

In this paper we consider the outflow of an initially cold
relativistic plasma and neglect only gravitation since  we are interested in
the effect of the electromagnetic forces
only  on the plasma.

\section{Basic equations and  methods of solution.}

The plasma outflow from astrophysical objects consists of two zones where
the properties of the flow differ qualitatively.
The nearest zone is located near the surface
of the central source and includes all the critical surfaces of the steady
state flow (Bogovalov \cite{bog97b}). The outer boundary of the nearest zone
can be located rather arbitrary but always downstream of the fast mode critical
 surface. The flow of the plasma in the nearest
zone is of the mixed type. It is subsonic near the source and is
super fast magnetosonic at the outer boundary.
The far zone is located downstream of the nearest zone. The plasma flow in the
far zone is super fast magnetosonic. Different methods 
should be applied for the solution of the problem in each of these zones.

The solution of the problem of the steady state plasma outflow
can be divided in two steps (Bogovalov \& Tsinganos \cite{bogts}).
In the first step, the  problem is solved in the
nearest zone by the relaxation method and in the second,
the steady state solution in the far zone is
obtained by solving directly the transfield equation as a Cauchy problem. 
The solution in the nearest zone is used to specify the initial values
for the flow in the far zone.

\subsection{Solution in the nearest zone.}

The system of equations  defining the dynamics of the relativistic
plasma in the ideal MHD approximation is as follows,
\begin{equation}
\rho c({\bf \dot U} +({\bf v}\cdot{\bf\nabla }){\bf U})
=q{\bf E} +{1\over c}{\bf j\times B},
\label{1}
\end{equation}
\begin{equation}
{\bf E} +{1\over c}{\bf v \times B}=0,
\label{2}
\end{equation}
\begin{equation}
{\bf \nabla\times B}={4\pi\over c} {\bf j} + {1\over c} \bf \dot E,
\label{3}
\end{equation}
\begin{equation}
{1\over c} \bf \dot B= -{\bf \nabla \times E},
\label{4}
\end{equation}
\begin{equation}
\dot \rho+({\bf\nabla \cdot  v}\rho)=0.
\label{4a}
\end{equation}

Here $q=({\bf \nabla \cdot \bf E})/4\pi$ is the induced electric charge  
density,
$\bf j$ is the
electric current density, $\bf E$ is the electric field,
${\bf U=v}\gamma/c$ is the
 spatial component of the plasma four-velocity and $\rho$ is the mass
 density of the plasma.

The problem was solved in the cylindrical system of coordinates. Some
details of the solution method are given in appendix A. The
numerical simulation was performed only in the quarter of the total box of
simulation due to the symmetry of the flow in relation to the
equatorial plane and the axis of rotation. The boundaries of the nearest
zone include  the surface of the star, the axis of rotation, the equator and
the outer boundaries, which are a cylinder (right hand side of the box)
 and a  plane perpendicular the axis of rotation at
some distance from the source (upper side of the box).
The  outer boundaries are located in the
supersonic domain of the flow beyond the fast mode critical surface.
This guarantees the independence of the solution in the nearest zone of the
position and conditions of the outer boundaries.

The number of  independent boundary conditions on the surface of the
star is equal to the number of MHD waves leaving the surface 
to satisfy the  causality in the flow (Bogovalov \cite{bog97b}).
They are as follows:\\
1. A constant plasma density  $\rho_0$.\\
2. A constant Lorentz-factor $\gamma_0$ of the plasma in the corotating
frame system.\\
3. A constant normal component of the poloidal magnetic field $B_0$.\\
4. The continuity of the tangential component of the poloidal
electric field.\\
5. A constant temperature of the plasma equal to zero.

The boundary conditions at the axis of rotation correspond to the axial
symmetry of the flow. The boundary conditions at the equator
 follow from the symmetry of the flow in  relation to the
equatorial plane. This means that $\rho(z)=\rho(-z)$, $v_r(z)=v_r(-z)$,
$v_z(z)=-v_z(-z)$,
$B_r (z)=B_r (-z)$, $B_z(z)=-B_z(-z)$, $v_{\varphi}(z)=v_{\varphi}(-z)$ and
$B_{\varphi}(z)=B_{\varphi}(-z)$.

At the outer boundaries only one side internal derivatives were used for
calculation of the derivatives in the finite differences, since the solution
does not depend on the conditions down the flow from these boundaries.

\subsection{Solution in the far zone}

The flow in the far zone is fast mode supersonic.
The system of equations describing the steady state flow is hyperbolic.
Therefore, a Cauchy problem can be formulated for this flow.

It is convenient to introduce the
flux function $\psi$ as follows

\begin{equation}
{\bf B}_{p}={{\bf\nabla}\psi\times \hat {\bf\varphi}\over r},
\label{psi}
\end{equation}
Where $B_p$ is the poloidal magnetic field of the axisymmetric outflow and
$\hat {\bf\varphi}$ is the unit azimuthal vector.

The stationary MHD equations admit well known four integrals
(Tsinganos \cite{tsin82}). They are:
\begin{description}
\item[($\alpha$)] The ratio of the poloidal magnetic and
mass fluxes,  $F(\psi )$
\begin{equation}
F (\psi ) = {B_{p}\over 4\pi \rho v_{p}c}
\,.
\label{f}
\end{equation}
\item[($\beta$)]  The total angular momentum per unit mass $L (\psi )$,
\begin{equation}
rU_{\varphi}-FrB_{\varphi}=L(\psi) 
\,.
\end{equation}
\item[($\gamma$)] The corotation frequency $\Omega (\psi )$ in the
frozen-in condition
\begin{equation}
U_{\varphi}B_{p} - U_{p}B_{\varphi} = r \gamma B_{p} \Omega (\psi )/c
\,.
\label{freez}
\end{equation}
\noindent
\item[($\delta$)] The total energy $W(\psi )$ per particle
in the equation for total energy conservation,
\begin{equation}
\gamma-F(\psi)r\Omega (\psi)B_{\varphi}/c=W(\psi)
\,.
\label{eq10}
\end{equation}
\end{description}
This system of equations should be supplemented by 
a relativistic relationship between 
the components of the four-velocity
\begin{equation}
\gamma ^{2}=1+U_{p}^{2}+U_{\varphi}^{2}.
\end{equation}

Momentum balance across the poloidal field lines is expressed by the
transfield equation.
For analysing the behavior of the plasma at large distances, it is
convenient to deal with this equation 
in an orthogonal curvilinear coordinate system ($\psi, \eta$) formed by
the poloidal magnetic field lines and by the lines of the poloidal
electric field. $\psi$ varies with the motion across the magnetic field lines,
while $\eta$ varies with the motion along the magnetic field lines.
A geometrical interval in these coordinates can be expressed as
\begin{equation}
(d{\bf r})^{2}=g^2_{\psi}d\psi^{2}+g^2_{\eta}d\eta^{2}+r^{2}d\varphi^{2},
\end{equation}
where $g_{\psi}, g_{\eta}$ are the corresponding line elements (components
of the metric tensor).

According to Landau \& Lifshitz (1975) the equation
$T_{\psi; k}^{k}=0$,
where $T^{ij}$ is the energy-momentum tensor and "; k" denotes covariant
differentiation, will have the following
form in these coordinates,
\begin{equation}
\begin{array}{c}
{\partial\over\partial\psi}\left[
{B^{2}-E^{2} \over 8\pi} \right] - {1\over r}{\partial r\over
\partial\psi}\left[U_{\varphi}v_{\varphi}c\rho-
{B_{\varphi}^{2}-E^{2} \over 4\pi} \right]-\\
\\
{1\over g_{\eta}} {\partial g_{\eta} \over \partial\psi}
\left[U_{p}v_{p}c\rho-{B_{p}^{2}-E^{2} \over 4\pi}\right] = 0
\end{array}
\label{transfield}
\end{equation}
with the  electric field $E={r\Omega\over c} B_p$.

The unknown variables here
are $z(\eta, \psi)$ and $r(\eta, \psi)$.
The metric coefficient $g_{\eta}$ can be obtained from the
transfield equation (\ref{transfield}),

\begin{equation}
 g_{\eta} =\exp{(\int\limits_0^\psi G(\eta,\psi)d\psi)}\,,
 \label{ga}
 \end{equation}
 where
        \begin{equation}
         G(\eta,\psi)={{\partial\over\partial\psi}\left[
       {B^{2}-E^{2}\over 8\pi}\right] - {1\over r} {\partial r\over
        \partial\psi} \left[U_{\varphi}v_{\varphi}c\rho-
      {B_{\varphi}^{2}-E^{2}\over 4\pi}\right]
      \over \left[ U_{p}v_{p}c\rho-{B_{p}^{2}-E^{2}\over 4\pi}\right]}\,.
          \label{G}
        \end{equation}
The lower limit of the integration in (\ref{ga}) is chosen to be 0
such that the coordinate $\eta$ is uniquely defined.
In this way $\eta$ coincides with the coordinate $z$ where the
surface of constant $\eta$ crosses the axis of rotation.

The metric coefficient $g_{\psi}$ can be obtained from definition (\ref{psi})
in terms of the
magnitude of the poloidal magnetic field  as follows
\begin{equation}
g_{\psi} ={1\over rB_{p} }
\,.
\end{equation}

The orthogonality condition
\begin{equation}
 r_{\eta}r_{\psi}+z_{\eta}z_{\psi}=0
\,,
\label{orto}
\end{equation}
and the relationships
\begin{equation}
g^2_{\eta}=r_{\eta}^{2}+z_{\eta}^{2},~~~~g^2_{\psi}=r_{\psi}^{2}+z_{\psi}^{2}
\label{galpha}
\end{equation}

give us that

\begin{equation}
r_{\eta}=-{z_{\psi} g_{\eta} \over g_{\psi}},~~~~z_{\eta}={r_{\psi} g_{\eta} 
\over g_{\psi}}
\,,
\label{za}
\end{equation}

\noindent
with $g_{\eta}$  calculated by the expression (\ref{ga}).
Here $r_{\eta}=\partial r/\partial\eta$, $z_{\eta}=
\partial z/\partial\eta$, $r_{\psi}=\partial r/\partial\psi$,
$z_{\psi}=\partial z/\partial\psi$.
For the numerical solution of the system of equations (\ref{za})
the two step Lax-Wendroff method on the lattice with a dimension equal
to 1000 is used.

Equations  (\ref{za}) should be supplemented by
appropriate boundary conditions and initial values on some initial surface
of constant $\eta$
located in the nearest zone, but down the flow from all the critical surfaces.
The form of the initial surface of constant $\eta$
was obtained 
numerically  through integration of the following equations,

\begin{equation}
{\partial r\over \partial\psi}={B_{z}\over rB_{p}^{2}},~~~~
{\partial z\over \partial\psi}=-{B_{r}\over rB_{p}^{2}}
\,.
\end{equation}

The values $B_z$, $B_r$ and integrals $W(\psi), L(\psi),
\Omega(\psi), F(\psi)$ were taken from the solution of
the problem in the nearest zone.

Boundary conditions on the axis of rotation and the equatorial plane are
the same as the conditions in the nearest zone. No conditions at  infinity
were specified.

\section{Parameterisation of the solution}

  A stationary solution of the relativistic plasma flow from the star with
an initially m-l magnetic field is defined by three
dimensionless parameters (Bogovalov \cite{bog99}). One of them is the ratio of 
the
radius of the star $R_{star}$ to the radius  of the initial fast mode
magnetosonic surface $R_f=\sqrt{(B_0R_{star})^2/4\pi\rho_0v_0cU_0}$,
where $U_0=v_0\gamma_0/c$.
The word "initial" here refers to the radius of the fast mode surface when the
star is not rotating.
But this dependence can be neglected under the condition
$R_{star} \ll R_f$ and $R_{star} \ll R_l$, where
$R_l=c/\Omega$ is the radius of the light cylinder, and $\Omega$ is the
angular velocity of the uniformly rotating star.
Therefore, the solution crucially
depends only on two dimensionless parameters which can be chosen rather
arbitrary. It is convenient to choose one of them as follows
\begin{equation}
\sigma=({R_f\over R_l})^2= {B_0^2\over 4\pi\rho_0\gamma_0v_0^2}
({R_{star}\Omega\over c})^2.
\label{10}
\end{equation}
$\sigma$ is the squared ratio of the initial radius of the fast mode
surface to the light cylinder radius. According to (\ref{10}) the
radius of the light cylinder in units of $R_f$ is
\begin{equation}
{R_l\over R_f}=\sqrt{1/\sigma}.
\label{11}
\end{equation}

The parameter $\sigma$ is closely related to the so-called magnetisation
parameter usually used  in the physics of the relativistic winds from radio
pulsars (Arons \cite{arons}). The magnetisation parameter is the ratio of the  
Poynting
flux density to the flux density of the mass energy 
(including rest mass energy).
The parameter $\sigma$ used here is equal to
the ratio of the Poynting flux to the
mass energy flux at the equator of the rotator with the initially
m-l magnetic field  provided that $\sigma/U_0^2 << 1$
(Bogovalov \cite{bog99}).
In our previous work (Bogovalov \cite{bog97a}),
we used another definition of the $\sigma$ parameter
introduced initially by Michel (\cite{michel}). This parameter
is denoted by 
$\sigma_M$.  There is a simple relationship  between them
$\sigma_M=\sigma U_0$. It follows from this relationship
that $\sigma_M$ is equal to the  Poynting flux density per  particle
on the equator  provided that $\sigma/U_0^2 << 1$.

It is convenient to take the second parameter as follows
\begin{equation}
\alpha={R_f\over R_l U_0}={\sqrt{\sigma}\over U_0}.
\label{alpha}
\end{equation}
The physical sense of this parameter becomes clear
if we pay attention to the fact that
in the model of Michel (\cite{michel}) the terminating four-velocity of the
plasma on
the equator is $U_{\infty}=(\sigma_M^{1/3})$ and $\alpha$ can be presented as
follows
\begin{equation}
\alpha=(U_{\infty}/U_0)^{3/2}.
\label{13}
\end{equation}
According to Eq. (\ref{13}), the parameter $\alpha$ shows how effectively the
magnetic forces arising in  the rotating magnetic field accelerate the
plasma. If
$\alpha < 1$ acceleration is not important compared to the initial
velocity of the plasma. It is natural to call the rotators with
$\alpha < 1$ as slow rotators. Correspondingly, the rotators with
$\alpha \ge 1$ should be fast
rotators which could in principle (but not necessarily) accelerate the
plasma. The parameter $\alpha$ introduced here is the relativistic
generalization of the $\alpha$ parameter introduced earlier by
Bogovalov \& Tsinganos (\cite{bogts}) for the nonrelativistic plasma.
Numerical
simulations of the nonrelativistic
plasma outflow show that this parameter indeed divides rotators into two
groups with different efficiency of acceleration and collimation of the
plasma. Below we will see  that this division remains valid also in the
relativistic limit.

Representation of the solution in the parameters $\alpha$, $\sigma$ is
convenient since the steady state outflow in the initially m-l
magnetic field has a simple analytical solution in the limit
$\alpha \rightarrow 0$ independent of the value of $\sigma$
(Bogovalov \cite{bog97a}) and the transition to the nonrelativistic
limit corresponds simply to
the limit $\sigma \rightarrow 0$ and $v_0/c \ll 1$.

For comparison with observation, Eq. (\ref{10}) should be expressed in
variables general for the m-l model  and real astrophysical objects.
An astrophysical object can be considered as equivalent to the
object with the m-l magnetic field if they both produce winds with
the same magnetic flux, the same  mass flux,
the same initial Lorentz factor and finally both rotate with the same
angular velocity.  The magnetisation parameter in these variables
takes the form
\begin{equation}
\sigma= {\psi_t^2\over (2\pi )^2m\dot N \gamma_0 v_0}({\Omega\over c})^2,
\end{equation}
where $\psi_t$ is the open field lines flux of one polarity in the wind,
$\dot N$ is the rate of ejection of particles with mass $m$.
For radio pulsars,  $\psi_t = \pi B_0 R_p^2$ where
$R_p =R_{star}\sqrt{R_{star}\Omega/c}$ is the radius of the polar
cap on the surface of the pulsar ( Ruderman \& Sutherland \cite{rs}). 

The position of the radio pulsars in the parameter space $\alpha$, $\sigma$
was discussed in detail by Bogovalov (\cite{bog99}). 
All radio pulsars eject Poynting-dominated winds ($\sigma > 1$) but basically
all of them are slow rotators ($\alpha < 1$). One of the youngest and
fastest rotating pulsars with a strong magnetic field on the surface
is the pulsar in the Crab Nebula. This pulsar rotates with
$\Omega \approx 200$ and ejects $e^{\pm}$ pairs with
the rate $\dot N \approx 4\cdot 10^{38}$ per second and average
$\gamma_0 \sim 200$ ( Daugherty \& Harding \cite{dh}). At
$B_0 \approx 4\cdot 10^{12}$ G the magnetisation parameter of the wind
$\sigma \approx 1.3 \cdot 10^4$ and $\alpha \approx 0.57$
The light cylinder radius of the Crab pulsar is located at the distance
$R_L =1.5\cdot 10^8$ cm from the axis of rotation. The radius of the
fast magnetosonic surface $R_f$  exceeds the
light cylinder radius  by a factor of 114. Marginally, the Crab  pulsar
could be the fast rotator provided that $\gamma_0 < 200$. The uncertainties
in estimates of this parameter do not rule out this possibility.
But all other,
more slowly rotating pulsars are certainly the slow rotators.

In AGNs and superluminal sources $\gamma_0$ is defined from observations of
jets and lies in the range $1-10$. If the collimation of the plasma
is really due to the SC process we should expect that $\alpha > 1$ and
$\sigma > \gamma_0 ^2$. So, the winds from  AGNs and superluminal sources
should be mainly Poynting flux dominated.

\section{Basic results}
\begin{figure}
\centerline{\psfig{file=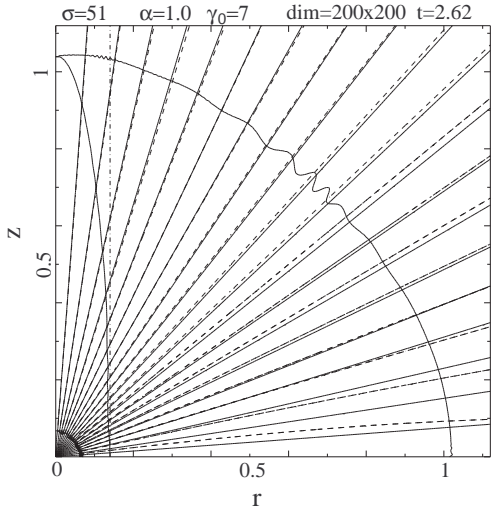,width=8.0truecm,angle=0}}
\caption{The steady-state relativistic plasma flow in the nearest zone
of the star for the parameters $\sigma$, $\alpha$, initial Lorentz-factor 
$\gamma_0$,
dimension of the lattice and duration of the simulation in units $R_f/v_0$ 
indicated
at the top of the figure. 
The cylindrical coordinates $z$ (axis of rotation) and $r$ (equator)
are given in units of $R_f$.
The star is located in the lower left-hand  corner of the figure. 
Solid lines radiating out from the star indicate the lines of
the poloidal magnetic field while dashed lines represent the lines of the poloidal
electric currents generated at the rotation.  Thick (thin) solid lines indicate 
the fast
(Alfvenic) mode surface, while the vertical dot-dashed line represents 
the light cylinder.}
\label{fig1}
\end{figure}

In Fig. \ref{fig1} we show the result of the numerical simulation of the
plasma
outflow in the nearest zone of the star, for a time interval $t=2.62 R_f/v_0$ 
and for the set of the values of the two key parameters, $\alpha =1.0$ 
and $\sigma=51$.
This time interval is larger by a factor of 2.62  compared to the typical 
time  the plasma and all MHD perturbations take to travel from the surface
of the star to  the fast mode surface. 
Note that during this time interval, a steady state solution is reached, 
\footnote{This solution is stable.
In the previous work (Bogovalov \cite{bog97a}) the time was
expressed in the units $1/\Omega$. Therefore, the ratio of the simulation
time in the present work over the  time in the previous work 
is $(t_{present}/t_{previous})(R_f\Omega/v_0) \approx
(t_{present}/t_{previous})\sqrt{\sigma}$ provided that $v_0\sim c$.
Here    $t_{present}$ and $t_{previous}$ are the dimensionless times in
the present and previous works.
In the present work the solution is stable up to the time $t_{present}=2.62$
at $\alpha =1$ and $\sigma=51$. In the previous work the solution for
$\alpha=0.0014$ and $\sigma=3.3$ was destroyed after $t_{previous}=2.02$.
Thus, in the present work much stronger Poynting-dominated wind
was simulated, for  9.26 times longer than in the previous one. The simulation
shown in the Fig. 4 lasted even longer. This confirms the stability of the
solution.}
since all perturbations which could be excited at the start of the simulation
have left the box of simulation. The initial fast mode surface is circular 
with a radius equal to 1 in our units. 
Note that the final fast mode surface shown in Fig. 1 has a small amplitude kink
structure at a polar angle around 45 degrees. Such a structure is formed
in all simulations performed with the present method and in the present geometry
(Washimi \& Shibata \cite{shibata}, Bogovalov \& Tsinganos \cite{bogts}).
Basically it is due to the shape of the surface of the star which is
actually approximated by a sequence of step functions in the cylindrical 
system of coordinates.

\begin{figure}
\epsfxsize=8.0 cm
\centerline{\epsfbox[0 0 491 491]{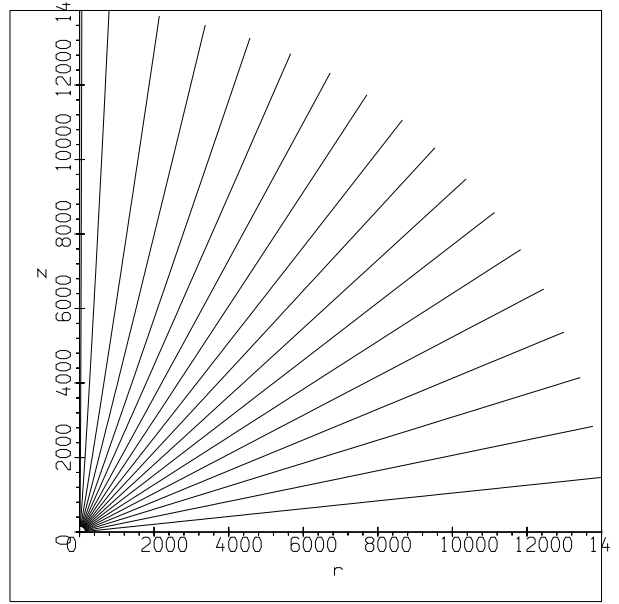}}
\caption{The relativistic plasma flow in the far zone for
$\alpha=1.0$ and $\sigma=51$. The coordinates are expressed in
 units of $R_f$.}
\label{fig2}
\end{figure}

The SC theory predicts that a wind of relativistic plasma consists 
of a cylindrically collimated core surrounded by a radially expanding flow 
(Heyvaerts \& Norman \cite{heyvaerts},
Chiueh et al. \cite{chiueh} and Bogovalov \cite{bog95}).
An inspection of Fig. \ref{fig1} shows that
the collimation of the plasma along the axis of rotation is practically
invisible in the nearest zone. The results of the same simulation in the far 
zone
are presented in Fig. \ref{fig2}. At a first glance, in the far zone
there is no cylindrical collimation.  
However, a closer examination of the results shows that a cylindrically 
collimated
core is actually formed in the wind.

\begin{figure}
\epsfxsize=8.0 cm
\centerline{\epsfbox[40 203 555 605]{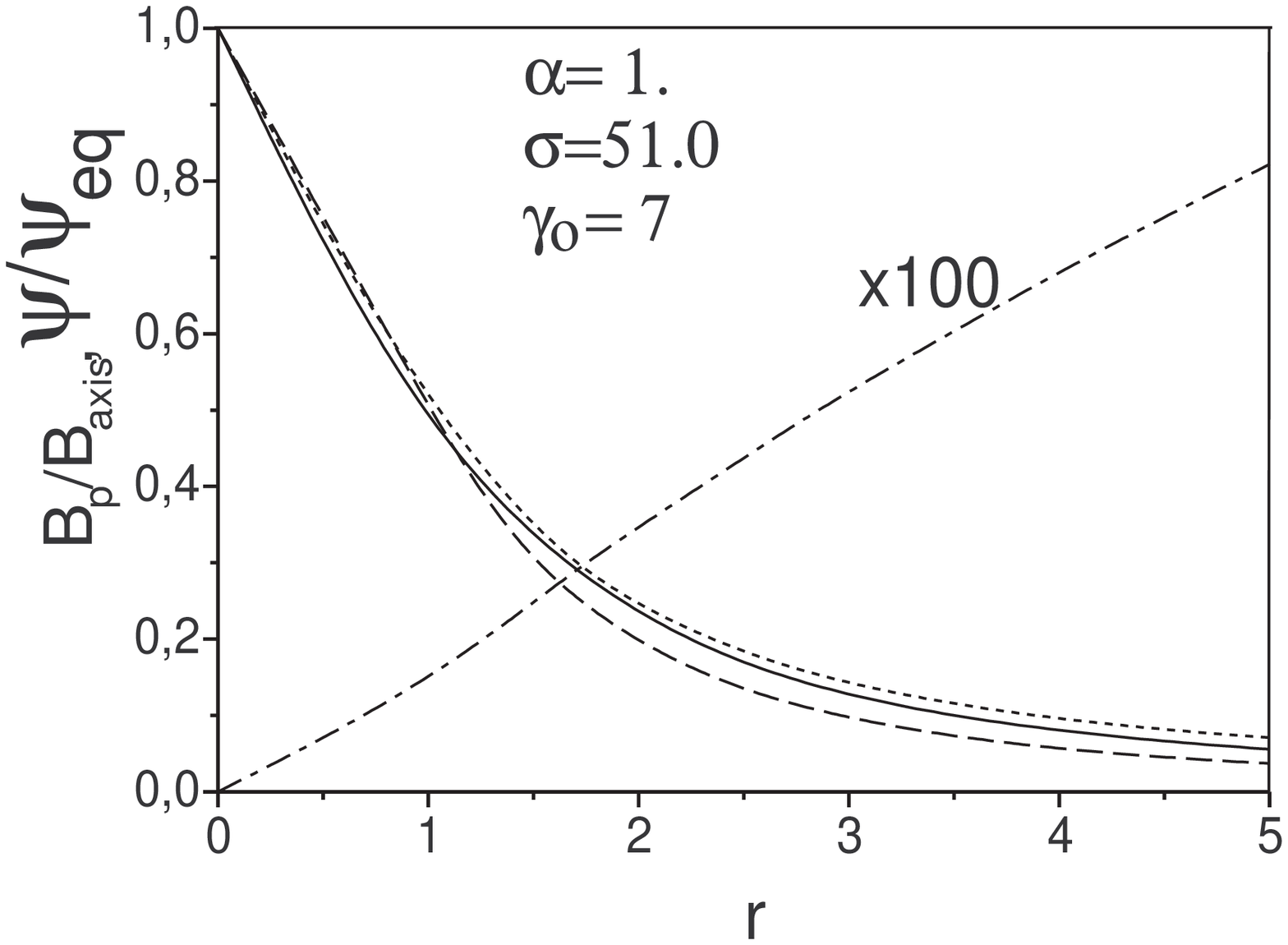}}
\caption{The distribution of the poloidal magnetic field across the jet
at $z=1121 R_f$ (solid line) and $z=140 R_f$ (dots), 
compared with the prediction of analytical theory
(dashed line). The poloidal magnetic field is normalized to its value
at the axis, $r=0$. The dashed-dotted line gives the distribution of the 
ratio of the magnetic flux to the total magnetic flux in the upper hemisphere
multiplied by 100. 
The distance to the axis of rotation $r$ is expressed in 
units of $R_f$ such that the radius of the jet is $1/\alpha=1$.}
\label{fig3}
\end{figure}

Fig. \ref{fig3} gives the distribution of the poloidal magnetic field
across the axis of rotation for two values of $z$. 
This distribution differs from that expected
for a m-l wind. The m-l poloidal magnetic field is described by the
equation $B_p=B_0/(r^2+z^2)$. And, at a fixed $z$ the poloidal field
normalized to its value at the axis of rotation ($B_{axis}$) depends on $r$
as  $B_p/B_{axis}=1/(1+(r/z)^2)$. Then, at $z=1121 R_f$ the poloidal magnetic 
field
remains practically constant at $r \le 5 \ll z=1121$.  This dependence
is represented by the horizontal line $B_p/B_{axis}=1$ in Fig. \ref{fig3}.
However, our numerical simulation gives a totally different picture.
Namely, the poloidal magnetic field has clearly a maximum at the rotation axis 
and then drops down with the distance $r$ from the axis, for example at $r \sim 
R_f$ 
to half its value from the axis.
It is important that this distribution is practically independent of  $z$. 
This result implies that the wind near the axis of rotation has achieved 
an asymptotic state and the flow is collimated exactly along the axis of 
rotation here.

The m-l model allows us to perform a quantitative comparison of the
SC theory and our numerical experiment.
The structure of the cylindrically collimated flow at large distances 
from the source is especially simple provided that the central source 
rotates uniformly while the plasma velocity and the ratio of the poloidal 
magnetic and mass fluxes (function $cF$ in  Eq. (\ref{f}))  
are constant across the jet.
In this case the dependence of $B_p$ on $r$ at the fixed $z$ for a 
cold plasma is given by the simple analytical expression
(Bogovalov \cite{bog95})
\begin{equation}
{B_p\over B_0}= {1\over 1+(r/R_j)^2},
\label{b}
\end{equation}
where $R_j=\gamma_jv_j/\Omega=R_f U_j/ U_0\alpha$ is the characteristic
radius of the collimated flow of the cold
plasma, while $\gamma_j$ and $v_j$ are the Lorentz factor and the velocity of 
the
plasma in this flow, respectively. Eq. (\ref{b}) is a direct consequence of the
SC theory and is valid at $r \gg R_l$.
The dependence of the normalised plasma density on $r$ is identical to the 
distribution (\ref{b})  since $F$ and the plasma velocity are assumed
to be constant across the cylindrically collimated flow. In this case
the plasma density is simply proportional to the poloidal magnetic field,
since $\rho=B_p/4\pi v_p cF(\psi)$ (see Eq. (\ref{f})).

All these conditions are indeed fulfilled in our case.
The outflow speed at the vicinity of the axis of rotation is close to the
speed of light ( $v_p=0.87c$ at $\gamma_0=2$ and $v_p=0.98c$ at
$\gamma_0=7$) and is independent of $r$. The function $F$ is assumed
independent of $\psi$ for all field lines and the star rotates uniformly.
A comparison of the calculated dependence of $B_p$ on $r$ with the
analytical prediction, Eq. (\ref{b}), shows that the SC theory correctly 
predicts the asymptotic properties of the wind structure.

The cylindrically collimated flow in Fig. (\ref{fig2}) is not visible 
since its transversal scale is extremely small. The flux of the
poloidal magnetic field contained in this flow is small as well. The
dependence of the magnetic flux on $r$ is shown in Fig. (\ref{fig3}) by
the dashed-dotted line.
For plotting convenience, this flux is multiplied by 100 and is also 
normalised to the total magnetic flux in the upper hemisphere, $\psi_{eq}$, 
which coincides with the magnetic flux at the equator.  
The mass flux normalised to the total mass flux
in the upper hemisphere is presented by the same curve since the
mass flux is proportional to the poloidal magnetic field in our case.
The fraction of the magnetic and mass flux in the cylindrically
collimated flow is of the order $0.1\%$.
The rest of the magnetic and mass flux expands radially, in accordance with
the SC theory.

\begin{figure}
\centerline{\psfig{file=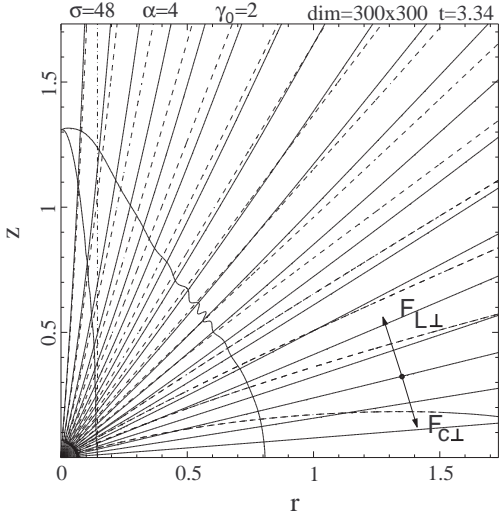,width=8.0truecm,angle=0}}
\caption{Steady-state relativistic plasma flow in the nearest zone
of the star and for the parameters $\sigma$, $\alpha$, initial Lorentz-factor 
$\gamma_0$,
dimension of the
lattice and the duration of the simulation in units $R_f/v_0$ indicated
at the top of the figure. Coordinates $z$ (axis of rotation) and
$r$ (equator) are given in units of $R_f$.
The star is located in the lower left-hand  corner of the figure. 
Solid lines radiating out from the star indicate the lines of
the poloidal magnetic field while dashed lines represent the lines of the poloidal
electric currents generated at the rotation.  Thick (thin) solid lines indicate 
the fast
(Alfvenic) mode surface, while the vertical dot-dashed line represents the light cylinder.}
\label{fig4}
\end{figure}

\begin{figure}
\epsfxsize=8.0 cm
\centerline{\epsfbox[0 0 491 491]{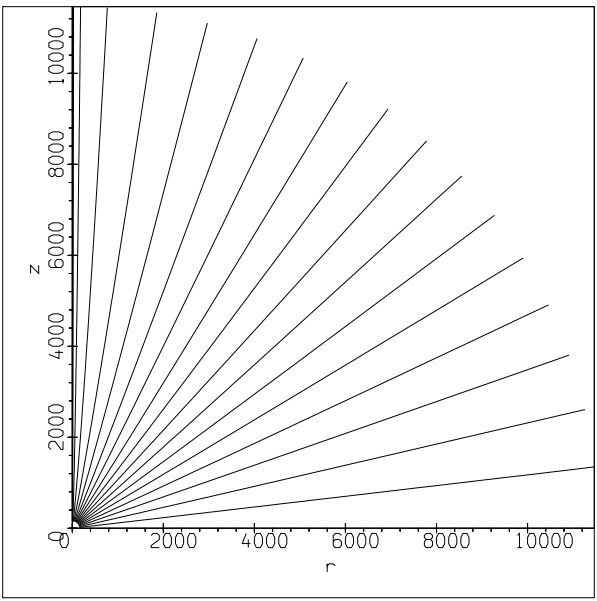}}
\caption{Relativistic outflow in the far zone of the star for
$\alpha=4$ and $\sigma=48$. The coordinates are expressed in  units
of $R_f$.}
\label{fig5}
\end{figure}

\begin{figure}
\epsfxsize=8.0 cm
\centerline{\epsfbox[40 202 552 596]{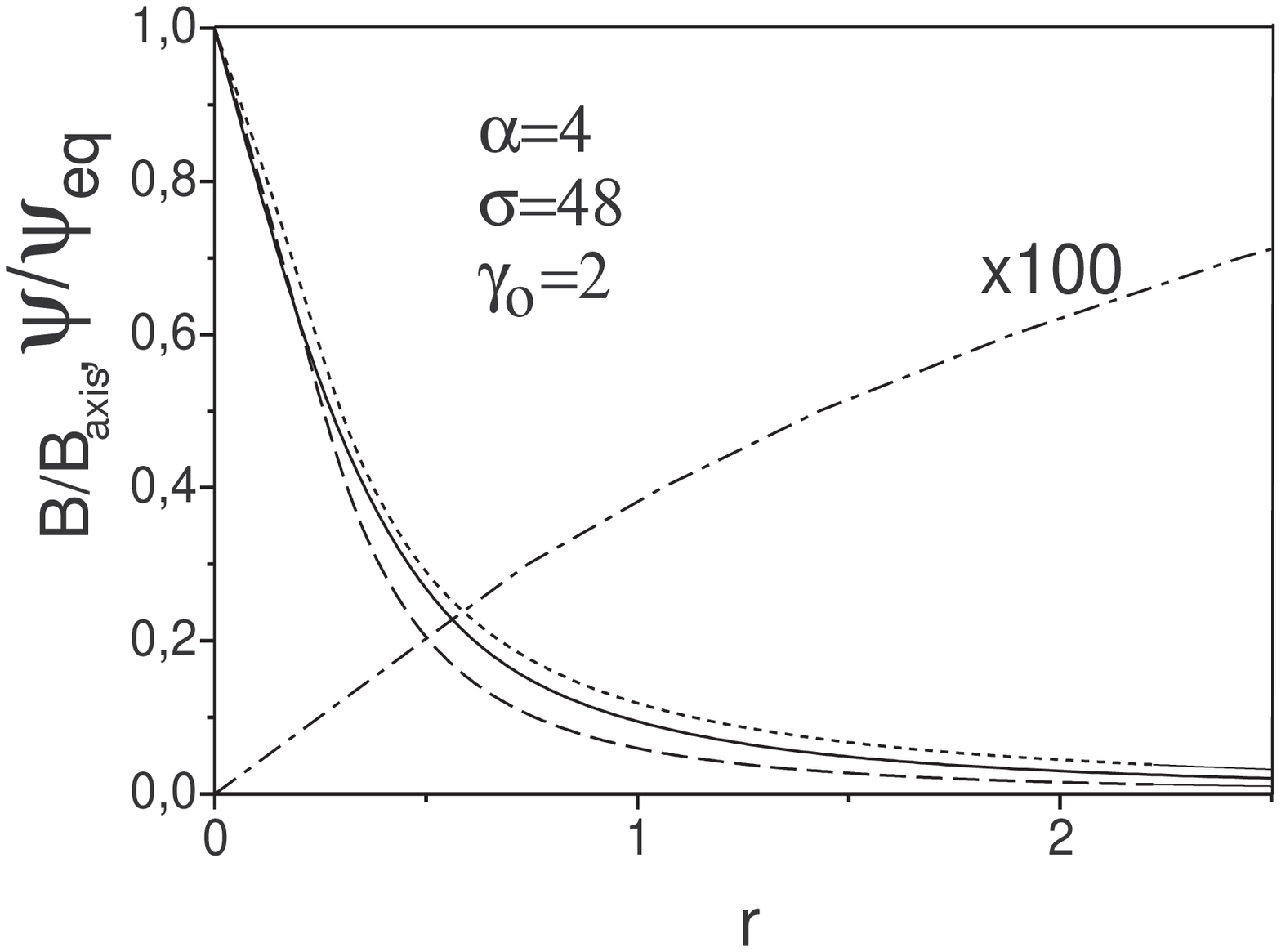}}
\caption{The dependence of the poloidal magnetic field across the jet
at $z=10350 R_f$ (solid line) and at $z=1000 R_f$ (dotted line)
compared with the prediction of analytical theory
(dashed line). The poloidal magnetic field is normalized to its value
at the axis, $r=0$. The dashed-dotted line gives the distribution of the 
ratio of the magnetic flux to the total magnetic flux in the upper hemisphere
multiplied by 100. 
The distance to the axis of rotation $r$ is expressed in 
units of $R_f$ such that the radius of the jet is $1/\alpha=0.25$.}
\label{fig6}
\end{figure}

The same physical picture emerges for a larger value of $\alpha$.
In Fig. (\ref{fig4}) we show in the nearest zone the results of 
the simulation of the flow of plasma for $\alpha=4$. 
The increase  $\alpha$ does  not  change significantly  the  structure  of  
the  flow,
although collimation now becomes   visible even in the nearest zone. The
same flow in the far zone is shown in Fig. (\ref{fig5}). It may be seen from 
this
figure that a very narrow collimated beam is formed in the wind. The
structure of the cylindrically collimated flow
is shown in Fig. (\ref{fig6}) for two values of $z$.

The efficiency of the acceleration of the plasma by a rotating object is 
important
in astrophysical applications. This acceleration in the model under 
consideration 
is expected to be most efficient near the equatorial plane since at the axis of 
rotation there is no accelerating force at all. 
The accelerating force in the poloidal direction
is proportional to the product $j_{\perp} B_{\varphi}$, where $j_{\perp}$ is
the component of the electric current perpendicular to a field line.
The accelerating force in the azimuthal direction is proportional to
the product $j_{\perp} B_p$. Thus, all accelerating forces are
proportional to $j_{\perp}$.  This implies that the acceleration occurs
only if the electric currents cross the poloidal field lines.  The
electric currents shown by the dashed line in Figs. (\ref{fig1}) and 
(\ref{fig4}) 
indeed cross the poloidal field lines. This is very well seen especially in Fig.
(\ref{fig4}) near the equator.

\begin{figure}
%\epsfxsize=8.0 cm
%\centerline{\epsfbox[29 26 1072 500]{gamma.eps}}
\centerline{\psfig{file=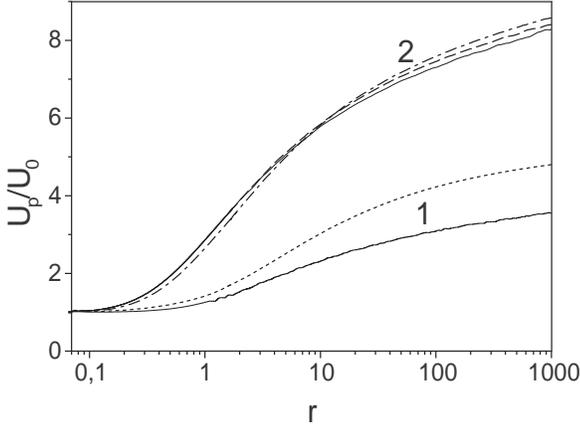,width=8.0truecm,angle=0}}
\caption{Dependence of $U_p$ along  a field line
on the distance to the centre of the source expressed in $R_f$.
Curve 1 is for $\alpha=1.0$, $\sigma=51$ for the equatorial plane.
Curves under number 2 are for the equatorial plane (solid line),
for $\psi/\psi_{eq}=0,7$ (dashed line), for $\psi/\psi_{eq}=0,3$
(dashed-dotted line) and for $\psi/\psi_{eq}=0,03$ (dotted line) for the
parameters  $\alpha=4$, $\sigma=48$.}
\label{fig7}
\end{figure}

The dependence of the four-velocity of the plasma on the distance from the 
centre 
of the source   is shown in Fig. (\ref{fig7}).
The distance is given in a logarithmic scale. The
energy of the particles increases slower than logarithmically and reaches 
a constant value. It follows from energy conservation (\ref{eq10})
that the maximum possible terminal four-velocity $U_{\infty}$
at the equator is $U_{\infty}=U_0 (v_{\infty}/v_0)\sigma$,
since $W$ at the equator can be
estimated as $\gamma_0\sigma$, provided that $\sigma \gg 1$ and the toroidal
magnetic field is estimated as $B_{\varphi} \approx - r\Omega B_p/v_0$
(Bogovalov \cite{bog97a}). All Poynting flux
should be transformed into kinetic energy of the plasma in this case.
But this value of the velocity is never achieved.
Michel (\cite{michel}) calculated for
$U_{\infty}$  the value
$U_{michel}=(\sigma U_0)^{1/3}$ for the plasma flow in a prescribed and fixed
m-l poloidal magnetic field. Therefore, 
$U_{michel}/U_0=\alpha^{2/3}$ (see Eq. \ref{13}).
It is reasonable to estimate the efficiency of the plasma acceleration
as the ratio $\xi=U_{\infty}/U_0 \sigma$ which is approximately 
the ratio of the part of
the Poynting flux transformed into the kinetic energy of the plasma (this is
valid, provided that $U_{\infty} \gg U_0$) to the
initial value of the Poynting flux at the equator in the relativistic limit.
For the Michel solution (\cite{michel})
$\xi=\alpha^{2/3}/\sigma$. The same value is given by  Beskin et al.
(\cite{beskin98}) for the analytical solution of
the same problem  under the  assumption that the poloidal
magnetic field goes to a monopole-like  field at $\alpha \rightarrow\infty$.
It follows from Fig. (\ref{fig7})  that $U_{\infty}/U_0$ increases with $\alpha$
essentially faster than $\alpha^{2/3}$. For the flow with $\alpha=4$  Michel's
estimate
gives $U_{p,\infty}/U_0=2.5$, while the numerical solution gives 
more than 8.
Correspondingly, the efficiency of the acceleration remarkably exceeds the
efficiency of the acceleration given by Michel and achieves  $16.7\%$ for
the flow at $\alpha=4$ and $\sigma=48$. It follows from Fig. \ref{fig7}
that the remarkable part of the acceleration occurs at large distances from
the star. It
happens at the collimation of the magnetic flux to the axis of rotation
due to  disbalance arising between the tension and pressure gradient of the
toroidal magnetic field (Begelman, \& Li \cite{li}, Fendt \& Camenzind
\cite{fendt96}). This mechanism was not taken
into account in the analysis of Michel (\cite{michel}) and Beskin et al.
(\cite{beskin98}) since
they assumed a fixed m-l magnetic field.

\begin{figure}
\centerline{\psfig{file=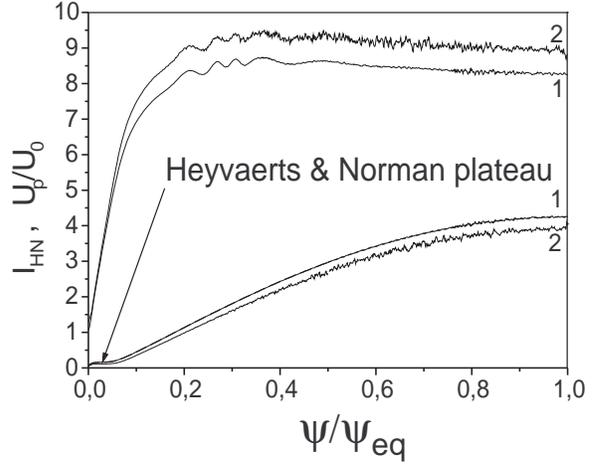,width=8.0truecm,angle=0}}
\caption{Dependence of the four-velocity of the plasma $U_p/U_0$
(upper curves) and
Heyvaerts \& Norman integral ${\rm I}_{HN}$ (lower curves) 
on $\psi$ for $z=1000R_f$ (curve 1) and
 $z=10350R_f$ (curve 2). The parameters of the outflow are  
$\alpha=4$, $\sigma=48$, $\gamma_0=2$.}
\label{fig8}
\end{figure}

It follows from Fig. \ref{fig7} that the largest energy is achieved not at the
equator but at $r=1000 R_f$ for $\alpha=4$. To understand why this happens it is
convenient to look on the dependence of $U_p/U_0$ on $\psi$
across field lines. This dependence is presented in Fig. (\ref{fig8}) by two
upper curves 1 ( for $z=1000R_f$) and 2 (for $z=10350R_f$). These curves
have small amplitude irregular structure which is purely of numerical origin.
It arises at large distances where the program deals with large
numbers. The maximum of the acceleration is achieved not at the equator,
but at $\psi/\psi_{eq} \sim 0.3$. This is explained by the fact that
the asymptotical regime predicted by the SC theory
is actually not achieved at these $z$ for $\psi/\psi_{eq} >0.05$.
It appears that the larger $\psi$
the larger the distance which is necessary to reach the asymptotic regime of the 
outflow.
This property of the SC process has already been  found
 in the nonrelativistic outflow simulation (Bogovalov \& Tsinganos 
\cite{bogts}). 

The physics of this phenomena can be understood as follows.
It can be shown from simple geometrical consideration that 

\begin{equation}
{1\over g_{\eta}} {\partial g_{\eta} \over \partial\psi} = {1\over
rR_{c}B_{p}}\,,
\end{equation}

\noindent
where $R_{c}$  is the curvature radius of the poloidal magnetic field
lines, with $R_{c}$ positive if the centre of curvature is in the domain
between the line and the axis of rotation and negative in the opposite case.
With this expression, Eq. (\ref{transfield}) takes the form

\begin{equation}
\begin{array}{c}
{\partial\over\partial\psi}\left[{B_{p}^{2}\over
8\pi}\right] +{1\over 8\pi r^{2}}{\partial\over
\partial \psi}\left[r^{2}(B^{2}_{\varphi}-E^{2})\right] - {1\over 4\pi r}
\left[{\partial r\over\partial\psi}\right]
{U_{\varphi}^{2}B_{p}\over U_{p}F(\psi)}\\
\\
-{[U_{p}-F(\psi)(1-(r\Omega/c)^{2})B_{p}]\over 4\pi rR_{c}F(\psi)}=0.
\end{array}
\label{treq}
\end{equation}

Let us consider the flow in the region of radial expansion, 
as in the case under consideration, where the SC is small. For simplicity, 
we neglect the acceleration of the plasma.  The first term in this
equation decreases with $r$ as $r^{-4}$. The second one decreases
as $r^{-2}$, since  $B_{\varphi} \sim E \sim rB_p$ (Bogovalov
\cite{bog97a}). The third term is proportional to the product
$rU_{\varphi}^2B_p$. This term decreases as $r^{-3}$, since $rB_p \sim 1/r$
and $U_{\varphi} \sim 1/r$ due to angular momentum conservation.
Thus,  only the
second and fourth terms survive in  Eq.  (\ref{treq})
at these large distances.
Therefore, this equation takes the form

\begin{equation}
{1\over 8\pi x^{2}}{\partial\over
\partial \psi}(x^{2}(B^{2}_{\varphi}-E^{2}))
-{(U_{p}+F(\psi)x^{2}B_{p})\over 4\pi xR_{c}F(\psi)}=0,
\label{trans4}
\end{equation}
where $x=r\Omega/c$.
From the frozen-in condition (\ref{2}) we may estimate the toroidal magnetic 
field at large distances when $v_{\varphi}\rightarrow 0$, where
 $v_{\phi}$ is the azimuthal velocity. 
\begin{equation}
B_{\varphi} =-{Ec\over v}.
\label{eqfr}
\end{equation}
In the asymptotic regime plasma moves along straight lines,
when $R_c \rightarrow \infty$. The structure of
the flow in this regime is defined by the trivial equation
\begin{equation}
{\partial\over\partial \psi}(x^{2}(B^{2}_{\varphi}-E^{2}))=0.
\end{equation}
Together with Eq. (\ref{eqfr}) this equation gives that the relativistic 
generalization
of the  Heyvaerts \& Norman integral $I_{HN}=|xB_{\varphi}|/\gamma$
(Chiueh et al. \cite{chiueh}) should be
constant in the region of radial expansion.  This integral, 
shown in Fig. (\ref{fig8}) by the two lower curves for essentially different
distances, is actually constant in the
small region $0.01 < \psi/\psi_{eq} < 0.05$ (Heyvaerts \& Norman
plateau in the figure). But at $\psi/\psi_{eq} > 0.05$, $I_{HN}$ is not
constant. This contradicts  the expected behavior of $I_{NH}$, but,
as we show below, this implies that the asymptotic regime is still not
reached in  this region of $\psi$.

The characteristic scale of collimation $D_{coll}$ which defines the distance
where the flow reaches the asymptotic regime can be estimated as follows.
$d\theta/dR=1/R_c$,  where $\theta$ is the
polar angle of the plasma velocity and
$R$ is the distance to the center of the source. In  the radially expanding
wind $d\psi= B_p r^2\sin(\theta)d\theta$. Taking into account also that at 
large distances $B_{\varphi} = - (r\Omega/v_p)B_p$ (Bogovalov \cite{bog97a}),
the rate of turning of the trajectory of the plasma follows from Eq.
(\ref{trans4}) and is defined by the
equation

\begin{equation}
|{d\theta\over dR}| \le {\sin{2\theta}\over RU_{0}^{2}({1\over\sigma}
+\sin^{2}{\theta})}.
\label{eq1}
\end{equation}
Integration of this equation gives the estimate of the distance at which the
line of the flow turns through  the angle $\Delta\theta$. The scale of the
collimation of the flow line $D_{coll}$ is defined by the distance
at which the flow line  turns through the angle $\Delta\theta \sim \theta$. The
integration of equation (\ref{eq1}) can be performed from the fast mode
surface  since this consideration is valid only in the supersonic region.
The scale of the collimation is estimated as
\begin{equation}
D_{coll}\ge R_{f}\exp
\left[ {\theta U_{0}^{2}({1\over\sigma}+\sin^{2}\theta)
\over \sin 2\theta} \right]
\,.
\label{eq2}
\end{equation}
According to this equation  there are two regions of the flow with the 
essentially different dependence of $D_{coll}$
on the parameters of the flow. The smallest  $D_{coll}$
takes place for small angles limited by the value
$\theta^2 < 1/\sigma$. For these angles the scale of the collimation is
(see also  Bogovalov \& Tsinganos \cite{bogts})
\begin{equation}
D_{coll}\ge R_{f}\exp{1\over 2\alpha^2}.
\label{eq3}
\end{equation}
Similar to the nonrelativistic outflow the collimation of the plasma
close to the rotational axis occurs near the fast mode surface at
$\alpha > 1$.
At the initially m-l outflow the magnetic flux depends on the angle
$\theta$ as $\psi=\psi_{eq}(1-\cos\theta)$. Therefore, the part of the
magnetic flux  where the collimation occurs as in the nonrelativistic
limit is 
\begin{equation}
{\psi\over\psi_{eq}} \le {1\over 2\sigma}.
\end{equation}
The collimation scale of the other part of the flow ( at low latitudes)
is defined by the equation
\begin{equation}
D_{coll} \ge R_f \exp{\theta\sin\theta U_0^2\over 2 \cos\theta}.
\label{26}
\end{equation}
and at the angles $\theta^2 \gg 2/U_0^2$ the collimation scale becomes
large compared with $R_f$ independent of $\sigma$.
It follows from this result that the remarkable bend of the flow lines
at low
latitudes is possible only for the plasma with relatively low $U_0$. The
collimation scale increases exponentially with the growth of the
Lorentz factor of the plasma. These properties of the flow
explain  the numerical results.

In the simulation we obtain such a behavior of the flow,  with three 
regimes: a core where the current increases with $\psi$, the Heyvaerts \& 
Norman regime and the  regime where the enclosed current increases 
(Fig. \ref{fig8}).  
Thus,  an extremely narrow cylindrical core  
is formed which contains the magnetic flux $\psi < (1/2\sigma)\psi_{eq}$. 
This core is formed at a distance comparable to $R_f$ for $\alpha >1$, 
similarly to the nonrelativistic limit. For larger $\psi$ the collimation
scale is defined by Eq. (\ref{26}). 
For $\gamma_0=7$ the characteristic scale
of collimation  is of the order of 
$R_f\cdot 10^{10.6\theta\tan\theta}$. It implies that the smaller $\theta$,
the earlier the asymptotic regime will be achieved. This explains why the
Heyvaerts \& Norman plateau is formed only at the relatively small $\theta$
(small $\psi$). At larger $\theta$ the scale of collimation exponentially
increases and finally diverges at $\theta = \pi/2$. Thus, the asymptotic
regime cannot be achieved at low latitudes. At the final distance from
the source there should exist a region at low latitudes (large $\psi$)
 where $I_{NH}$ is not constant. It is this behavior which is found in our
 calculations.

For $\alpha =4$ the collimation is more efficient and the effect of
collimation appears already in the nearest zone where the plasma is
not yet accelerated.
At the estimation of the collimation scale 
it is necessary to take into account acceleration of the plasma
to a Lorentz factor greater than 10 near the equator.
This could be done by integration of Eq. (\ref{eq1})
from a radius $R_{in} > R_f$ where the energy of the plasma practically
reaches the terminal value.
If we take into account the acceleration, the asymptotic
regime (in which $I_{NH}$ does not depend on $\psi$ in a radially expanding 
wind) will be achieved at a distance
of the order $D_{coll} > R_f 10^{22\cdot\theta\tan\theta}$. 
It follows from this expression that at $\alpha=4$ the asymptotic regime
at $R\approx 10^4\cdot R_f$ will be achieved only for $\psi/\psi_{eq} < 0.1$.
This value agrees well  with
the numerical results presented in Fig. \ref{fig8} in the limits of uncertainties.

Comparison of the present work with our previous work
(Bogovalov \& Tsinganos \cite{bogts} ) shows that the relativistic
plasma is collimated less efficiently than the nonrelativistic one.
In particular, the fraction of the collimated mass flux in the nonrelativistic
flow was
close to $1\%$, while in the  relativistic limit it is of the order $0.1\%$.
This difference can be understood from simple qualitative consideration.
The dynamics of the cold plasma is basically controlled by two forces. 
One of them is the volumetric Lorenz force,  
$({\bf j}\times {\bf B})/c$ and the other is the volumetric Coulomb force, 
$q\bf E$ (see Eq.(1)). The collimation of the plasma to
the axis of rotation is defined by the components of these forces
perpendicular to the stream lines. The perpendicular component of the
Lorenz force is $F_{L\perp}={1\over c}j_{\parallel} B_{\varphi}$,
where  $j_{\parallel}$ is the
component of the poloidal electric  current parallel to the
poloidal magnetic field and $B_{\varphi}$ is the toroidal magnetic field.
The perpendicular component of the Coulomb force is 
$F_{C\perp}=q\cdot E_{\perp}$. These forces are shown schematically
in Fig. (\ref{fig4}). It is important to note that these two forces are directed 
along opposite  directions. The Lorenz force collimates the plasma while the
Coulomb force decollimates it. In the nonrelativistic limit the
Coulomb force is neglected, since it is negligible in comparison to the Lorenz 
force.
The ratio of the Coulomb and Lorenz forces is of the
order $(v/c)^2$ in this limit. In the relativistic limit the Coulomb force
becomes practically  equal to the Lorenz force and it
strongly decreases the SC of the plasma. That is why the relativistic
generalization of the
parameter $\alpha$ introduced earlier for the nonrelativistic winds
(Bogovalov  \& Tsinganos \cite{bogts}) has additionally
the Lorentz factor in the denominator (see Eq. (\ref{alpha})).

Although the Lorentz factor in the denominator of Eq. (\ref{alpha})
decreases remarkably the value of $\alpha$,
it was expected that the collimation of the relativistic plasma will be
strong enough at the condition $\alpha > 1$. 
In the nonrelativistic case the collimation becomes
strong already in the subsonic region  under this condition.
The most surprising
result is that the relativistic plasma is practically
not collimated,  even under this condition. The increase of the
collimating force does not provide  more efficient collimation of the
relativistic plasma! This happens since another
relativistic effect comes into play.

The curvature  $\kappa_c = 1/R_c$ of the stream lines of the nonrelativistic
plasma is defined by the simple equation
\begin{equation}
\kappa_c = {F_{\perp} \over mnv^2},
\label{nr}
\end{equation}
 where $m$ is mass of particles, $F_{\perp}$ is the sum of all forces
affecting the plasma perpendicular
to the stream line. It is seen from this equation that an increase of the
force $F_{\perp}$ results in an increase of $\kappa_c$, such that collimation 
becomes stronger. In other words, as it is evident from 
Eq. (\ref{nr}), $\kappa_c$ is proportional to the
collimating force and inversly proportional to the  
mass density of the outflowing plasma. 
The situation changes essentially in the case of the relativistic, 
Poynting dominated plasma ($\sigma >1$) where the energy flux is
dominated by the electromagnetic field. Then, according to the familiar 
relationship
between energy and mass (${\it E}= mc^2$), the flux of the energy
is equivalent to some inertial mass flux. 
It follows from Eq.(\ref{trans4}) that eq. (\ref{nr})
for the relativistic Poynting-dominated plasma is modified as follows
\begin{equation}
\kappa_c = {F_{\perp} \over nv^2\gamma(m+ {B_{\varphi}^2\over 4\pi nc^2})}.
\end{equation}
This equation demonstrates that the effective mass of the Poynting-
dominated plasma is  dominated by the toroidal magnetic field. It is
important that now the effective mass increases by a term proportional to 
$B_{\varphi}^2$.
The collimating force is also proportional to the same term
(see eq. (\ref{trans4})). Therefore, the increase of the collimating force
is accompanied by a corresponding increase of the
effective mass density of the plasma. This in turn results in the fact that the
curvature $\kappa_c$ of the stream lines  does not increase with
the increase of the collimating  force. This is the reason why the collimation
scale (\ref{26}) at  low latitudes does not depend on the magnetic field and
angular velocity of the source.

\section{Conclusion}
 This work shows that the  self-collimation (SC)  process
occurs in relativistic winds as 
predicted by the SC theory. The wind consists 
of a radially expanding part which surrounds a 
cylindrically collimated flow.
The radius of the cylindrically collimated flow is $R_j \approx R_f/\alpha$,
 where $R_f$ is the initial radius of the fast magnetosonic surface.
The mass flux in the collimated part 
is small compared with the total mass flux in the wind ($\sim 10^{-3}$), 
even for relatively small Lorentz factors. SC essentially
decreases for larger Lorentz factors.  Acceleration of the
plasma occurs  in the  radially expanding wind while 
the plasma in the jet is not accelerated in this cold outflow model.

The relatively small part of the mass flux in the cylindrically collimated flow
has been also found in our simulations of nonrelativistic
outflows (Bogovalov \& Tsinganos \cite{bogts}), although the magnetic flux 
contained in the
relativistic case appears smaller by a factor of order 10. 
In addition, the collimation scale at low latitudes
does not depend on the magnetic field and the angular velocity
of the central source for $\alpha > 1$, in contrast with our results for
the nonrelativistic winds in the same model.  This means
that the weak collimation of the relativistic  plasma is due to the
intrinsic property of the plasma and not to a property of the model.
Collimation decreases due to the affect of the electric force which
becomes comparable with the collimating Lorenz force in the relativistic
 
limit and in the high $\sigma$ limit the collimation is saturated due to the 
contribution of the electromagnetic field into the inertial mass of plasma.

The most striking feature of the collimated flows obtained in the model
under consideration is that a very small fraction ($\sim 10^{-3}$)
of the mass  flux exists in the collimated flow in comparison to the mass flux
in the wind. This result is apparently in conflict with observations of AGN.
The source of the wind energy is the gravitational energy released by the
accreted matter. Therefore, the energy flux in the wind should be less than
the luminosity of the
accretion disk where about half of the gravitational energy is released
(Shakura \& Sunyaev \cite{ss}). The total luminosity of AGN $L_{tot}$ in
soft electromagnetic emission ( below X-rays) should be
comparable with the energy released at the accretion. According to our
results the energy flux in the jet $\dot E_{jet}$ should be by the factor
$10^{-3}$ less
than the accretion energy, $\dot E_{jet} < 10^{-3}\cdot L_{tot}$.
This disagrees with observations.  X-ray and
gamma-ray emission is usually attributed to the emission from jets
(Urry \& Padovani \cite{urry}).
Sometimes, the luminosity of the jets  is less only by  a factor of 10 than 
the total luminosity of 
AGN at other wavelengths (Elvis et al. \cite{elvis}).
The energy flux in this emission can be considered as the lower limit on the
energy flux in the jet (however, we do not take into account beaming here).
If so, we have in AGN
$\dot E_{jet} > 10^{-1}\cdot L_{total}$.

This disagreement between theory and observations implies that
either the present model has no observed counterpart or
the SC mechanism is not able in principle to provide
collimation of the major fraction of the outflow into the jet.
It is important in this regard to understand from the theoretical
point of view what modifications should be made in order to provide
magnetic collimation of the major fraction of the relativistic wind. 
This problem has been already investigated by 
Bogovalov \& Tsinganos (\cite{bogts00}) in relation to 
nonrelativistic winds.  It was shown in this study that in a  
model consisting of a central source ejecting a high-speed wind
and an accretion disk ejecting a nonrelativistic  magnetised outflow,   
all the mass flux from the  central source can be cylindrically collimated 
indirectly by the surrounding disk-wind.
Apparently, a collimation of the major part of a relativistic outflow from a
central
source can be provided by the same mechanism in the context of the
present model as well.

\begin{acknowledgements}
I acknowledge support from the University Joseph Fourier and the Institut
Universitaire de France as well as the warm hospitality of
Prof. G. Pelletier and Dr. J. Ferreira during my stay in the Laboratoire
d'Astrophysique de l'Observatoire de Grenoble,
where part of this work was done.
This work was partially supported by the Ministry of Education of Russia in
the framework of the program ``Universities of Russia - fundamental research'',
project N 990479 and by collaborative INTAS-ESA grant N 99-120. The author is
sincerely grateful to the unknown second referee and K.Tsinganos
for the large amount of work they put in
reading  of this manuscript and their addition of
useful comments.
The author thanks V.B. Komberg for useful discussion of the observational
properties of AGN.
\end{acknowledgements}

\appendix
\section{Details of the numerical simulation in the nearest zone.}
For the numerical
solution of the time dependent problem,  the system of equations
should be presented as
\begin{equation}
{\partial\over\partial t}\hat \Phi = A(\hat\Phi, \hat\Phi_i, \hat\Phi_{ik}),
\label{5}
\end{equation}
where $\hat\Phi$ is the vector of the plasma state which includes
the density, components of the velocities and magnetic fields.
$\hat\Phi_i$ and $\hat\Phi_{ik}$ are the first and the second order
derivatives of $\hat\Phi$ on the spatial
variables. Eq. (\ref{1}) contains
the time derivative $\bf\dot E$ in the right hand part generated by the
displacement current in Eq. (\ref{3}). Therefore, some special
transformation of Eq. (\ref{1}) is necessary to get form (\ref{5}).
The specific form of this transformation is not unique. In our previous
work (Bogovalov \cite{bog97a}) unsuccessful choice of this transformation
resulted in the instability of the numerical code in the 
Poynting flux-dominated regime. Here we use another form of the equation of
motion which provides stable numerical code in all regimes.

This equation can be obtained as follows.
Time derivative of the  frozen-in condition (\ref{2}) gives that
\begin{equation}
{\bf\dot E}= -{1\over \gamma}({\bf\dot U\times B})-
{1\over \gamma}({\bf U\times \dot B}) -{\dot\gamma\over\gamma}
{\bf E}.
\label{6}
\end{equation}
Substitution of Eq. (\ref{6}) and Eq. (\ref{3}) together with the relativistic
relationship $\dot\gamma=v_i\dot U_i/c$ in Eq. (\ref{1})
gives the equation of motion in the form
\begin{eqnarray}
\lefteqn{M_{ik}\dot U_k= -\rho({\bf v\cdot\nabla})U_i+qE_i/c+
{1\over 4\pi c}(({\bf \nabla\times B)\times B})_i}\nonumber\\
&&-{1\over 4\pi c^3}
(v_i({\bf B\cdot \dot B})-\dot B_i({\bf B\cdot v})),
\label{7}
\end{eqnarray}
where
$$
M_{ik}=(\rho+{B^2\over 4\pi c^2\gamma})\delta_{ik}-{1\over 4\pi c^2\gamma}
(B_iB_k+S_iv_k/c)
$$
and $S_i=({\bf E\times B})_i$. Terms with $\bf\dot B$ in the right hand part of
Eq. (\ref{7}) are eliminated with help of Eq. (\ref{4}). After that,
multiplication of equation of motion (\ref{7}) on the inverse matrix
\begin{eqnarray}
\lefteqn{M_{ik}^{-1}={1\over (\rho + {B^2\over 4\pi c^2 \gamma})}
[\delta_{ik}+}\nonumber\\
&&{B_iB_k\over 4\pi\rho c^2\gamma}+
{({\bf v\cdot B}/c)S_i B_k/4\pi\rho c^2\gamma +S_i v_k
/c\over (4\pi\rho c^2\gamma+B^2-({\bf S \cdot v}/c)}]
\end{eqnarray}
gives the equation of motion in the form (\ref{5}) appropriate
for the numerical solution.

A two-step Lax-Wendroff numerical scheme
was used for the numerical solution of the problem (Press et al. 
\cite{press}).
The analytical solution
obtained for the slow rotators (Bogovalov \cite{bog97a}) was used as the
initial state of the plasma.

\begin{figure}
\centerline{\psfig{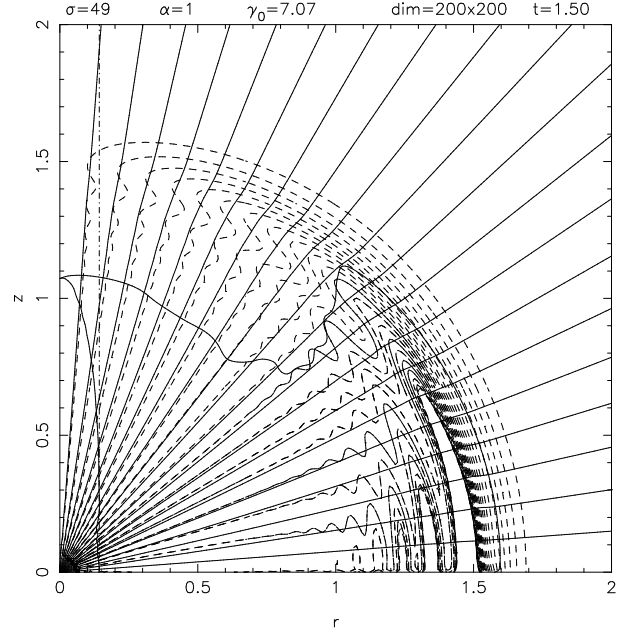}}
\caption{The simulation of propagation of the switch on shock wave in the
relativistic wind. All the notations are the same as in Fig.
\protect\ref{fig1} and \protect\ref{fig4}.
The wave is generated at the beginning of rotation of
the central object.}
\label{test}
\end{figure}

The testing of the numerical code was performed by comparison of the
results with the results obtained with the nonrelativistic version of the
code (Bogovalov \& Tsinganos \cite{bogts}), by comparison with predictions
of the analytical theory (Bogovalov \cite{bog92}, Bogovalov \cite{bog97a})
and by  simulation of the perturbation propagation in the wind.
One of these tests is presented in Fig. \ref{test}. 
The beginning of the star rotation  starts the
shock wave propagating with the Alfvenic velocity in the
isotropic wind. This velocity
is very close to 1 for the taken parameters of the wind. 
The shock has  an oscillating structure typical for the collisionless
shocks. The front is smoothed due to the numerical diffusion.
The first (largest) maximum can be accepted as the boundary of the shock.
It follows from the figure that the wave propagates with the velocity
very close to the light velocity which is equal
$c/v_0= 1.01$  in the dimensionless variables normalized to $v_0$
at $\gamma_0=7$.

\end{document}